\documentclass[]{article}
\usepackage{amsmath,amssymb}
\title{Fitting a Sum of Exponentials to Numerical Data}
\author{Bernhard Kaufmann\footnote{Metallurgical research department, voestalpine, 4031 Linz,
Austria. \newline e-mail: bernhard.kaufmann@voestalpine.com}}
\date{}

\begin{document}
\maketitle
\begin{abstract}
A finite sum of exponential functions may be expressed by a linear
combination of powers of the independent variable and by
successive integrals of the sum. This is proved for the general
case and the connection between the parameters in the sum and the
coefficients in the linear combination is highlighted. The fitting
of exponential functions to a given data- set is therefore reduced
to a multilinear approximation procedure. The results of this
approximation do not only provide the necessary information to
compute the factors in the exponents and the weights of the
exponential terms but also they are used to estimate the errors in
the factors.
\end{abstract}

\section{Introduction}
From time to time the need arises to fit a sum of exponentials to
numerical data. That means to approximate a given data- set
consisting of pairs of real numbers $(x_j, y_j)$ by the following
expression:
\begin{equation} \label{expsum}
y(x) = a_0 + \sum_{i=1}^N a_i e^{-b_i x}
\end{equation}
where \(x, x_j \in \mathbb{R^+}, y_j \in \mathbb{R} \), \(N \in
\mathbb{N^+} \) and $a_i$, $b_i$ are unknown real numbers which
have to be chosen so that the fit becomes optimal.

If the $b_i$ were known, the task usually would be a well posed
linear problem, but if the $b_i$ are unknown too, it turns out to
be ill conditioned. The hopelessness of efforts dealing with this
kind of problem has been described drastically by F.S. Acton
\cite{ac} in a chapter entitled "What not to compute".

At first sight, fitting equation \ref{expsum} to a given data- set
inevitably seems to be a nonlinear problem. However it has been
noted \cite{squire}, \cite{diamessis} that equation \ref{expsum}
may be expressed as a linear combination of powers of x and
successive integrals of y(x), reducing the problem to a
multilinear fitting procedure. This method is based on the fact
that y(x) can be shown to satisfy an ordinary linear differential
equation of N-th order with constant coefficients. The roots of
the characteristic polynomial of this equation give the $b_i$ and
the $a_i$ are identified as solutions to linear equations
involving the $b_i$ and the derivatives of y(x) at x = 0. However,
derivatives of experimental data- sets enhance the errors in the
data, therefore it is desirable to eliminate them. Actually,
\cite{diamessis} shows already the way to do this, but only for
the case N = 2 and $a_0 = 0$, the general case not really being
obvious. In the present paper, this method of linearizing the
fitting procedure is revived and without referring to differential
equations and without using derivatives, the general case is
derived. Additionally, a method for estimating the errors in the
exponential factors is presented. Of course, the problem remains
ill posed, but linear fitting offers computational advantages over
nonlinear approximation and also supplies estimates of the errors
in the computed coefficients, which may be used to predict the
errors in the exponential coefficients.

\section{Results}
The functions used to construct a linear approximation problem are
powers of x and successive integrals of y(x). Before the announced
relation can be asserted, some definitions are required.

\textbf{Definition:} The k-{th} integral of y(x) is defined
recursively:
\begin{eqnarray}
I_0(x) & = & y(x) \nonumber \\
I_k(x) & = & \int_{0}^{x} I_{k-1}(t)dt \qquad k > 0 \label{defI}
\end{eqnarray}

\textbf{Definition:} $\beta_{Nij}, \alpha_{Nij}$. Given the set \(
B = \{b_1,\cdots,b_N\} \) of N exponential factors $b_k$ in
equation \ref{expsum} we consider the products of i different
elements of $B$. Each of these products corresponds to a
combination of i elements out of $B$, the number of these products
therefore is
\begin{equation}
C(N,i) = \frac{N!}{(N-i)!i!} = \binom{N}{i}
\end{equation}
as is proved in combinatorics. We assume that the products are
ordered in some way. $\beta_{Nij}$ then is the j-{th} of these
products. Additionally, we define $\beta_{N01}$ = 1.

$\alpha_{Nij}$ is the sum of all $a_k$ in equation \ref{expsum}
excluding those whose index is equal to that of one of the b's in
$\beta_{Nij}$. By definition, $\alpha_{N01}=\sum_{k=0}^N a_k$.
Obviously, each $\alpha_{Nij}$ contains at least $a_0$.
\\
\textbf{Example:} N=3

\begin{displaymath}
\begin{array}{lll}
\beta_{301}=1  \\
\beta_{311}=b_1 & \beta_{312}=b_2  & \beta_{313} = b_3 \\
\beta_{321}=b_1 b_2  & \beta_{322}=b_1 b_3 & \beta_{323}=b_2 b_3   \\
\beta_{331}=b_1 b_2 b_3 \\ \\
\alpha_{301}=a_0+a_1+a_2+a_3  \\
\alpha_{311}=a_0+a_2+a_3 & \alpha_{312}=a_0+a_1+a_3
& \alpha_{313}=a_0+a_1+a_2 \\
\alpha_{321}=a_0+a_3 & \alpha_{322}=a_0+a_2 &
\alpha_{323}=a_0+a_1 \\
\alpha_{331}=a_0 \\

\end{array}
\end{displaymath}

With these definitions, the central statement of this article now
may be asserted:

\begin{equation} \label{thatsit}
a_0 + \sum_{i=1}^N a_i e^{-b_i x} = -\sum_{i=1}^N I_i(x)
\sum_{j=1}^{C(N,i)}\beta_{Nij} +\sum_{i=0}^N \frac{x^i}{i!}
\sum_{j=1}^{C(N,i)} \beta_{Nij} \alpha_{Nij}
\end{equation}

Assuming the validity of equation \ref{thatsit} the task now
consists in approximating the data- set $\{(x_j, y_j)\}$ by a
linear combination of the 2N functions $(I_1,\hdots,I_N,$ $x,
\hdots, x^N)$ plus a constant. By standard linear approximation
techniques the coefficients $(c_1, \hdots, c_N,$ $d_1,\hdots,
d_N)$ and the intercept $d_0$ may be determined together with
their errors $(\Delta c_1, \hdots, \Delta c_N,$ $\Delta
d_1,\hdots, \Delta d_N)$ and $\Delta d_0$. It follows that

\begin{eqnarray}
c_i & = & -\sum_{j=1}^{C(N,i)}\beta_{Nij} \qquad i=1,\hdots, N
\label{Vieta}\\
d_i & = & \sum_{j=1}^{C(N,i)} \frac{\beta_{Nij}\alpha_{Nij}}{i!}
\qquad i=1,\hdots, N \label{a1} \\
d_0& = & \alpha_{N01}=\sum_{k=0}^N{a_k} \label{a2}
\end{eqnarray}

Given the $c_i$ in \ref{Vieta} Vieta's root theorem asserts that
the $b_i$ are the N roots of the polynomial
\begin{equation} \label{polynom}
P(x) = x^N+\sum_{i=1}^N(-1)^{i+1}c_ix^{N-i}
\end{equation}
As soon as the $b_i$ are known, the expressions \ref{a1} and
\ref{a2} represent a system of N+1 linear equations for the N+1
coefficients $a_i$.

If the $\Delta c_i$ are small, the relation between the errors may
be approximated by the linear terms of the Taylor- series for
P(x).
\begin{equation}
\Delta P(x) = \frac{\partial P(x)}{\partial x} \Delta x +
\sum_{i=1}^N \frac{\partial P(x)}{\partial c_i}\Delta c_i
\end{equation}

As the $b_k$ are roots of P, \(\Delta P\) should be zero and
therefore, inserting $b_k$ for x, we get:
\begin{equation} \label{xerror}
\Delta b_k = -\frac{1}{\frac{\partial P(b_k)}{\partial
x}}\sum_{i=1}^N \frac{\partial P(b_k)}{\partial c_i}\Delta c_i
\end{equation}
Treating the $c_i$ and $b_k$ as probability variables with
standard deviations $s_{ci}$ and $s_{bk}$, the standard deviation
and therefore the estimated error of $b_k$ is given by
\begin{equation}
s_{bk} = \frac{1}{|\frac{\partial P(b_k)}{\partial
x}|}{\sqrt{\sum_{i=1}^N b_k^{2(N-i)}s_{ci}^2 +
2\sum_{i=1}^N\sum_{j=i+1}^N(-1)^{i+j}b_k^{2N-i-j}Cov(c_i, c_j)}}
\end{equation}

As usual, $Cov(c_i, c_j)$ means the covariance between $c_i$ and
$c_j$.

It remains to show that equation \ref{thatsit} is valid. For this
purpose it is useful to state some properties of the coefficients
$\beta$.

For any $l$ with $1 \le l < N$ and any $i \le N$ the sum of all
$\beta_{Nlm}$ may be divided into the sum of all $\beta_{Nlm}$
containing $b_i$ and those not containing $b_i$:
\begin{equation} \label{indgen}
\sum_{m=1}^{C(N,l)}\beta_{Nlm} ~=~
b_i\sum_{j=1}^{C(N-1,l-1)}\beta_{(N-1)(l-1)j}^{(-i)}+\sum_{j=1}^{C(N-1,l)}\beta_{N-1,l,j}^{(-i)}.
\end{equation}
With $\beta_{(N-1)lj}^{(-i)}$ we denote the products not
containing $b_i$ that is, which are chosen from the set $\{b1,
\hdots, b_{i-1}, b_{i+1}, \hdots, b_N\}$ containing N-1 elements
and not containing $b_i$. For $l \le 0$ we define
$\beta_{(N-1)lm}^{(-i)}=1$.

An important special case of \ref{indgen} results if i = N. Then
$\beta_{(N-1)lm}^{(-N)}$ = $\beta_{(N-1)lm}$ and the following
expression results:

\begin{equation} \label{ind1}
\sum_{m=1}^{C(N,l)}\beta_{Nljm}=\sum_{j=1}^{C(N-1,l)}\beta_{(N-1)lj}+b_{N}\sum_{j=1}^{C(N-1,l-1)}\beta_{(N-1)(l-1)j}
\end{equation}

For a proof of equation \ref{thatsit} consider the following
system of equations:

\begin{eqnarray}
I_0(x) & = & a_0 + \sum_{i=1}^N a_i
e^{-b_i x} \nonumber \\
I_k(x) & = & a_0 \frac{x^k}{k!} + \sum_{i=1}^N a_i \left[
\left(\frac{-1}{b_i}\right)^{k} e^{-b_i x}  -
\sum_{j=0}^{k-1}\left(\frac{-1}{b_i}\right)^{k-j}
\frac{x^j}{j!}\right] \label{defEq}
\end{eqnarray}

The validity of \ref{defEq} is easily seen by performing the
integrals in equation \ref{defI} analytically.

Now consider the following linear transformations defined
recursively on the set of equations \ref{defEq}:

\begin{eqnarray}
I_k^{(1)} & = & I_k + b_1 I_{k+1} \nonumber \\
I_k^{(h)} &=& I_k^{(h-1)} + b_h I_{k+1}^{(h-1)} \qquad h > 1
\label{defTransf}
\end{eqnarray}

For this kind of transformation a rather general relationship
holds:

\begin{equation} \label{lemTransf}
I_k^{(h)} = I_k + \sum_{l=1}^h \sum_{m=1}^{C(h,l)} \beta_{hlm}
I_{k+l}
\end{equation}

\textbf{Proof:} Induction for h. For h = 1, proposition
\ref{lemTransf} just repeats the definition of $I_k^{(1)}$. Now
assume that \ref{lemTransf} holds for $I_k^{(h)}$. Then

\begin{displaymath}
I_k^{(h+1)} = I_k^{(h)}+b_{h+1} I_{k+1}^{(h)} =
\end{displaymath}
\begin{displaymath}
I_k + \sum_{l=1}^h \sum_{m=1}^{C(h,l)} \beta_{hlm} I_{k+l} +
b_{h+1}I_{k+1}+ \sum_{l=1}^h  \sum_{m=1}^{C(h,l)}
b_{h+1}\beta_{hlm} I_{k+l+1} =
\end{displaymath}
\begin{displaymath}
I_k + \sum_{l=1}^h \sum_{m=1}^{C(h,l)} \beta_{hlm} I_{k+l} +
b_{h+1}I_{k+1}+ \sum_{l=2}^{h+1}  \sum_{m=1}^{C(h,l-1)}
b_{h+1}\beta_{h(l-1)m} I_{k+l} =
\end{displaymath}
\begin{displaymath}
I_k + \sum_{m=1}^{C(h+1,1)} \beta_{(h+1)1m} I_{k+1} +
\sum_{l=2}^{h} \sum_{m=1}^{C(h+1,l)} \beta_{(h+1)lm} I_{k+l} +
b_{h+1} \beta_{hh1}I_{k+h+1}
\end{displaymath}
where \ref{ind1} has been used in order to obtain the last line.
Obviously, this result may be converted into
\begin{displaymath}
I_k^{(h+1)} = I_k + \sum_{l=1}^{h+1} \sum_{m=1}^{C(h+1,l)}
\beta_{(h+1)lm} I_{k+l}
\end{displaymath}
whereby the proof of \ref{lemTransf} is completed.

Consider now $I_0^{(N)}$. By \ref{lemTransf}
\begin{equation}
I_0^{(N)} = y + \sum_{l=1}^N \sum_{m=1}^{C(N,l)} \beta_{Nlm}I_l
\end{equation}
Inserting \ref{defEq} this expands into
\begin{displaymath}
y + \sum_{l=1}^N \sum_{m=1}^{C(N,l)} \beta_{Nlm}I_l =
\end{displaymath}
\begin{displaymath}
a_0+\sum_{i=1}^N a_i e^{-b_i x} \left[1+\sum_{l=1}^N
\sum_{m=1}^{C(N,l)}\beta_{Nlm}\left(\frac{-1}{b_i}\right)^l
\right]+
\end{displaymath}
\begin{equation} \label{basis}
\sum_{l=1}^N \sum_{m=1}^{C(N,l)}\beta_{Nlm}\left[a_0
\frac{x^l}{l!}-\sum_{i=1}^N a_i
\sum_{j=0}^{l-1}(-b_i)^{j-l}\frac{x^j}{j!}\right]
\end{equation}

The motive for applying transform \ref{defTransf} to $I_0$ was to
get rid of the exponential terms. The following proposition
asserts that equation \ref{basis} is actually free of exponential
terms:

\begin{equation} \label{lemMinOne}
\sum_{l=1}^h \sum_{m=1}^{C(h,l)} \beta_{hlm} \left(\frac{-1}{b_i}
\right)^l = -1 \qquad 1 \leq i \leq h
\end{equation}
\textbf{Proof:}

For h = 1 the assertion is trivial. Now assume that
\ref{lemMinOne} is valid for h. Then the following calculations
prove the truth for h+1 and therefore for all h:
\begin{displaymath}
\sum_{l=1}^{h+1} \sum_{m=1}^{C(h+1,l)} \beta_{(h+1)lm}
\left(\frac{-1}{b_i} \right)^l =
\end{displaymath}

\begin{displaymath}
\sum_{l=1}^h \left(\frac{-1}{b_i} \right)^l
\left[\sum_{m=1}^{C(h,l)} \beta_{hlm} + b_{h+1}
\sum_{m=1}^{C(h,l-1)} \beta_{h(l-1)m} \right] +
\left(\frac{-1}{b_i} \right)^{h+1} \beta_{(h+1)(h+1)1} =
\end{displaymath}
\begin{displaymath}
-1
-\frac{b_{h+1}}{b_i}+b_{h+1}\sum_{l=1}^{h-1}\sum_{m=1}^{C(h,l)}\beta_{hlm}\left(\frac{-1}{b_i}
\right)^{l+1}+\left(\frac{-1}{b_i}
\right)^{h+1}\beta_{(h+1)(h+1)1} =
\end{displaymath}
\begin{displaymath}
-1-\frac{b_{h+1}}{b_i}-\frac{b_{h+1}}{b_i}\sum_{l=1}^{h}\sum_{m=1}^{C(h,l)}\beta_{hlm}\left(\frac{-1}{b_i}
\right)^{l} + \frac{b_{h+1}}{b_i} \left(\frac{-1}{b_i}
\right)^{h}\beta_{hh1}+
\end{displaymath}
\begin{displaymath}
\left(\frac{-1}{b_i} \right)^{h+1}\beta_{(h+1)(h+1)1}
\end{displaymath}
Using equation \ref{lemMinOne} for and collecting all terms the
last expression evaluates to -1.

To complete the proof of equation \ref{thatsit} some more
transformations on formula \ref{basis} are required:

\begin{displaymath}
y + \sum_{l=1}^N \sum_{m=1}^{C(N,l)} \beta_{Nlm}I_l =
a_0+\sum_{l=1}^N \sum_{m=1}^{C(N,l)}\beta_{Nlm}\left[a_0
\frac{x^l}{l!}-\sum_{i=1}^N a_i
\sum_{j=0}^{l-1}(-b_i)^{j-l}\frac{x^j}{j!}\right]=
\end{displaymath}
\begin{displaymath}
\sum_{i=0}^N a_i+\sum_{l=1}^N
\frac{x^l}{l!}\left[\sum_{m=1}^{C(N,l)} \beta_{Nlm}
a_0-\sum_{j=l+1}^N \sum_{m=1}^{C(N,j)} \beta_{Njm}\sum_{i=1}^N
a_i(-b_i)^{l-j}\right]=
\end{displaymath}
\begin{displaymath}
\sum_{i=0}^N a_i+\sum_{l=1}^N
\frac{x^l}{l!}\left[\sum_{m=1}^{C(N,l)} \beta_{Nlm}
a_0+\sum_{i=1}^N
a_i(-b_i)^{l}+\sum_{j=1}^l\sum_{m=1}^{C(N,j)}\beta_{Njm}\sum_{i=1}^N
a_i (-b_i)^{l-j}\right]=
\end{displaymath}
\begin{equation} \label{intermed}
\sum_{i=0}^N a_i+\sum_{l=1}^N
\frac{x^l}{l!}\left[\sum_{m=1}^{C(N,l)} \beta_{Nlm} \sum_{i=0}^N
a_i+\sum_{i=1}^N a_i S_{il} \right]
\end{equation}
where
\begin{displaymath}
S_{il}=
\sum_{p=1}^{l-1}\sum_{m=1}^{C(N,l-p)}\beta_{N(l-p)m}(-b_i)^{p}+
(-b_i)^l \qquad \textnormal{for }l > 1
\end{displaymath}
and
\begin{displaymath}
S_{i1}= -b_i    \qquad \textnormal{for }l = 1
\end{displaymath}

Using \ref{indgen}, for $l > 1$ $S_{il}$ transforms into
\begin{displaymath}
S_{il} =
\end{displaymath}
\begin{displaymath}
\sum_{p=1}^{l-1}\left(\sum_{m=1}^{C(N-1,l-p)}\beta_{(N-1)(l-p)m}^{(-i)}+b_i
\sum_{m=1}^{C(N-1,l-p-1)}\beta_{(N-1)(l-p-1)m}^{(-i)}
\right)(-b_i)^p+(-b_i)^l
\end{displaymath}
Substituting in the second part of this sum q for p+1 this
expression transforms into
\begin{displaymath}
S_{il} =
-\sum_{m=1}^{C(N-1,l-1)}\beta_{(N-1)(l-1)m}^{(-i)}b_i+\sum_{p=2}^{l-1}\sum_{m=1}^{C(N-1,l-p)}\beta_{(N-1)(l-p)m}^{(-i)}(-b_i)^p
\end{displaymath}
\begin{displaymath}
-\sum_{q=2}^{l-1}\sum_{m=1}^{C(N-1,l-q)}\beta_{(N-1)(l-q)m}^{(-i)}(-b_i)^q
-(-b_i)^l+(-b_i)^l =
\end{displaymath}
\begin{displaymath}
-\sum_{m=1}^{C(N-1,l-1)}\beta_{(N-1)(l-1)m}^{(-i)}b_i
\end{displaymath}
Therefore $S_{il}$ is the negative sum of all $\beta_{Nlm}$ which
contain $b_i$. Consequently, in expression \ref{intermed} only
those $a_i$ are not cancelled for which $\beta_{Nlm}$ does not
contain $b_i$. For that, \ref{intermed} may be written as
\begin{displaymath}
y + \sum_{l=1}^N \sum_{m=1}^{C(N,l)} \beta_{Nlm}I_l =\sum_{l=0}^N
\frac{x^l}{l!} \sum_{m=1}^{C(N,l)} \beta_{Nlm} \alpha_{Nlm}
\end{displaymath}
which proves equation \ref{thatsit}.
\\
\textbf{Example:} Consider this sum of two exponentials and a
constant:
\begin{displaymath}
y(x) = 0.3 + exp(-0.7 x)+0.4 exp(-0.3 x)
\end{displaymath}
The function is evaluated in in the interval $0 \le x \le 6$ at Np
equally spaced points. The discrete function values are multiplied
by one plus a gaussian distributed random variable so that the
relative error has the standard deviation $\sigma$. I1 and I2 are
calculated using the trapezoidal method. For different settings of
Np and $\sigma$ the coefficients c1 and c2, d0 (the intercept), d1
and d2 and the corresponding errors of c1 and c2 as well as the
covariance between these two factors are determined by the
commercial statistics program
STATISTICA\textsuperscript{\circledR} and are listed in table 1.
The parameters b1, b2, a0, a1, a2 and the errors $\Delta b_1$ and
$\Delta b_2$ are calculated from these coefficients as described
above and are listed in table 2.

\begin{table}[ht]\caption{Statistically determined coefficients}
\footnotesize
\begin{tabular}{|r|r|l|l|l|l|l|l|l|l|l|}
\hline
Id & Np & $\sigma$ & $c_1$ & $\Delta c_1$ &$c_2$ & $\Delta c_2$ & Cov &$d_0$ & $d_1$ & $d_2$\\
\hline
1 & 601 & 0.0000 & -1.0000 & $10^{-6}$ & -0.2100 & $10^{-6}$ & 0 &1.7000 & 0.8800 & 0.0315\\
2 & 601 & 0.0001 & -1.0054 & 0.0024 & -0.2130 & 0.0014 & $3.10^{-6}$ &1.7001 & 0.8890 & 0.0320\\
3 & 601 & 0.001  & -1.0358 & 0.0253 & -0.2300 & 0.0146 & 0.000343 &1.7008 & 0.9396 & 0.0349\\
4 & 601 & 0.01   & -1.0842 & 0.2954 & -0.2615 & 0.1700 & 0.0502 &1.7006 & 1.0249 & 0.0408\\
5 &2001 & 0.01   & -0.9155 & 0.1190 & -0.1603 & 0.0685 & 0.00815 &1.6999 & 0.7369 & 0.0228\\
\hline
\end{tabular}
\end{table}

\begin{table}[ht] \caption{Parameter estimates based on the coefficients
in table 1} \footnotesize
\begin{tabular}{|r|l|l|l|l|l|l|l|}
\hline
Id & $b_1$ & $\Delta b_1$ & $b_2$ & $\Delta b_2$ & $a_0$ & $a_1$ & $a_2$\\
\hline
1 & 0.7000 & $3.10^{-6}$ & 0.3000 & $3.10^{-6}$ & 0.30 & 1.00 & 0.40 \\
2 & 0.7020 & 0.0019 & 0.3034 & 0.0021 & 0.30 & 0.99 & 0.41 \\
3 & 0.7134 & 0.0180 & 0.3224 & 0.0196 & 0.30 & 0.95 & 0.45 \\
4 & 0.7220 & 0.1211 & 0.3622 & 0.1754 & 0.31 & 0.88 & 0.51 \\
5 & 0.6796 & 0.0281 & 0.2359 & 0.0911 & 0.28 & 1.09 & 0.32 \\
\hline
\end{tabular}
\end{table}

The results show that for small errors in the coefficients the
estimated variance of $b_1$ and $b_2$ is also small and the
estimate is realistic. The first case was computed without
artificial noise, in this case the accuracy seems to be determined
mainly by the statistics program. Adding noise deteriorates the
accuracy of the results rapidly. While a relative error of 0.0001
(case 2) still leads to a reasonable result, the tenfold relative
error (case 3) already means that the calculated uncertainty of
$b_2$ is about 7\%. A one- percent inaccuracy in the data (case 4)
gives a result even with the first digit uncertain. As case 5
where the number of data- points is raised to 2001 shows,
increasing the size of the data- set may at least partially
compensate for noise.

\end{document}